\documentclass[aps,prl,nofootinbib,preprintnumbers,amsmath,amssymb,latexsym,array,enumerate,letter,twocolumn,superscriptaddress]{revtex4}
\usepackage{amssymb}
\usepackage{amsmath}
\usepackage{epsfig}
\usepackage{hyperref}
\usepackage{breakurl}
\usepackage{xcolor}

\makeatletter
\def\simgt{\mathrel{\lower2.5pt\vbox{\lineskip=0pt\baselineskip=0pt
           \hbox{$>$}\hbox{$\sim$}}}}
\def\simlt{\mathrel{\lower2.5pt\vbox{\lineskip=0pt\baselineskip=0pt
           \hbox{$<$}\hbox{$\sim$}}}}
\makeatother

\newcommand{\be}{\begin{equation}}
\newcommand{\ee}{\end{equation}}
\newcommand{\bea}{\begin{eqnarray}}
\newcommand{\eea}{\end{eqnarray}}
\newcommand{\beq}{\begin{eqnarray}}
\newcommand{\eeq}{\end{eqnarray}}

\def\lsim{\mathrel{\rlap{\lower4pt\hbox{\hskip1pt$\sim$}}
     \raise1pt\hbox{$<$}}}         
\def\gsim{\mathrel{\rlap{\lower4pt\hbox{\hskip1pt$\sim$}}
     \raise1pt\hbox{$>$}}}         

\begin{document}

\preprint{LCTP-18-04}

\title{Searching for Dark Photon Dark Matter with Gravitational Wave Detectors }

\author{Aaron Pierce}
\affiliation{Leinweber Center for Theoretical Physics, Department of Physics, University of
Michigan, Ann Arbor, MI 48109}

\author{Keith Riles}
\affiliation{Department of Physics, University of Michigan, Ann Arbor,
MI 48109}

\author{Yue Zhao}
\affiliation{Leinweber Center for Theoretical Physics, Department of Physics, University of
Michigan, Ann Arbor, MI 48109}

\begin{abstract}
If dark matter stems from the background of a very light gauge
boson, this gauge boson could exert forces on test masses in
gravitational wave detectors, resulting in  displacements with a
characteristic frequency set by the gauge boson mass.  We outline a
novel search strategy to hunt for such dark matter, and show that
both ground-based and future space-based gravitational wave
detectors have the capability to make a 5$\sigma$ discovery in
unexplored parameter regimes.
\end{abstract}

\maketitle

\section{Introduction}

Dark matter (DM) makes up the dominant form of matter in the
universe, but the properties of the particles that compose it
remain unknown. If the DM particle is a boson, it can be extremely
light, with masses bounded by limits from dwarf galaxy morphology,
$m  \gsim  10^{-22}$ eV, see, e.g. \cite{Chen:2016unw}.\footnote{We
use natural units $c=\hbar=\sqrt{4\pi \varepsilon_0}=1$, but provide
in the Appendix a translation of a critical expression to SI units.}   A gauge
boson, here denoted ``dark photon"  (DP) can be naturally light and
is a candidate for the DM.
The relic abundance may be produced by a
misalignment mechanism associated with the inflationary epoch, as first
discussed in \cite{Nelson:2011sf}, with additional discussion and
resolution of subtleties in \cite{Arias:2012az,Graham:2015rva}.
Other non-thermal production mechanisms are possible
\cite{Co:2017mop}.  A generic feature of these production mechanisms is that the DPDM remains decoupled from the thermal bath, and so it effectively cools to be non-relativistic before matter-radiation equality and acts as cold dark matter.

When the dark matter is very light, its local occupation number is
much larger than one. It  can then be treated as a coherently oscillating
background field with oscillation frequency determined by its
mass.  DPDM therefore imparts external oscillating forces acting on
objects carrying non-zero dark charge.  While the identity of the DPDM is
model dependent, we will consider gauged baryon number, $U(1)_B$, and baryon number minus lepton number, $U(1)_{B-L}$, as
benchmarks in later discussions.

The strongest constraints on the coupling of light gauge bosons to
the Standard Model (SM) come from equivalence principle tests,
including those from the E$\ddot{\textrm{o}}$t-Wash group
\cite{Su:1994gu,Schlamminger:2007ht} and Lunar Laser Ranging (LLR)
experiments \cite{Williams:2004qba,Talmadge:1988qz,Turyshev:2006gm}.
In such experiments, the Earth provides a large dark charge,
sourcing a dark photon field.\footnote{These constraints could in principle be
evaded in non-minimal scenarios wherein the Earth captured
background particles charged under the dark gauge group, thereby
screening the charge.}
Another potential powerful constraint comes from consideration of
black hole superradiance, initially proposed in
\cite{Arvanitaki:2009fg} to probe spin-0 particles, such as the axion.
It was generalized in
\cite{Baryakhtar:2017ngi,East:2017ovw,East:2017mrj,Cardoso:2018tly}
for a light vector boson.  The absence of related signals could rule out some of the regions of interest discussed here.  Importantly, however, the
effective superradiance requires the absence of
non-gravitational interactions to a very precise degree
\cite{Baryakhtar:2017ngi,Arvanitaki:2014wva}. Self-interactions can
be easily introduced in a dark sector for a massive gauge boson.
Indeed, depending on the DPDM production mechanism, such
interactions may be expected.  We do not discuss these bounds
further.

Recent detections of gravitational waves (GW) by the LIGO and Virgo
detectors, see e.g.~\cite{GW150914,GW170814,GW170817}, have opened
the era of GW astronomy. These interferometers currently measure
strain amplitudes of transient GW signals at better than 10$^{-21}$,
with improvements of a factor of $\sim$3 expected in the next
several years to reach design sensitivities~\cite{Aasi:2013wya}.
These strain measurements hinge on sub-attometer sensitivities to
the relative displacement of mirrors located 3-4 km apart. As we
will discuss, relative displacements of the test masses
(interferometer mirrors) may be generated not only by the passage of
GW, but also by a DPDM background.  A somewhat related idea of using
GW detectors to search for clumps of dark matter via the induced
displacements has been discussed in
\cite{Adams:2004pk,Hall:2016usm}. For longer-lived signals,
integration over long observation times can yield strain
sensitivities orders of magnitude lower than is possible for
transient signals. The DM galactic velocity dispersion is
$v_{0}\sim\mathcal{O} (10^{-3}) $, thus the coherence time is
$\sim\mathcal{O}(10^6)$ oscillation periods ($10^6/f$).

In this letter, we propose a novel search which can be carried out
by GW detectors, presently with Laser Interferometer Gravitational-Wave Observatory (LIGO)/Virgo and in the future with the Laser Interferometer Space Antenna  (LISA).  Both
 ground-based and space-based experiments have the potential to probe an unexplored parameter space of DPDM.

\section{Dark Photon Dark Matter Induced Relative Displacement}

Owing to its light mass, DPDM will be coherent over long length
scales.  Its spatial coherence length can be estimated as
$\ell_{coherence} = 2\pi/(m_{A} v_{0})$, where $m_A$ is the mass of
the dark photon, and $v_{0}$ corresponds to a typical dark matter
virial velocity in the halo, $v_{0} \sim 10^{-3}$.

For frequencies corresponding to near the best sensitivity of LIGO, $m_A
= 2\pi f = 2\pi (100 \, \textrm{Hz})=4\times 10^{-13}$ eV, and
$v_{0}=10^{-3}$, we have $\ell_{coherence}\simeq 3\times 10^6$ km.

The local amplitude $A_{\mu,0}$  of the dark gauge field $A_{\mu}$
can be found by equating its energy density, $\frac{1}{2} m_{A}^2
A_{\mu,0}A_{0}^{\mu}$, to that of the local dark matter, for which
we take a fiducial value of  $\rho_{DM}$=0.4 GeV/cm$^{3}$. Within a
coherence length, $A_\mu(t,\vec x) \simeq A_\mu(t)\simeq A_{\mu,0}
\sin (m_A t - \vec{k}\cdot \vec{x}).$ This oscillating dark photon
field will act as an external force on the test objects of GW
detectors, and the resulting displacements may be detected by such
experiments.

Since the DPDM is non-relativistic, the electric
components associated with the time derivative of the field are
much larger than the magnetic components.
The
acceleration acting on a test mass located at $x_{i}$ is
\begin{eqnarray}\label{eq:acc}
\vec{a}_i(t,\vec {x_i})&=&\frac{\vec{F}_i(t,\vec {x_i})}{M_i}\simeq\epsilon e
\frac{q_{D,i}}{M_i}
\partial_t \vec{A}(t,\vec {x_i})\nonumber\\
&=& \epsilon e \frac{q_{D,i}}{M_i} m_A \vec A_0 \cos{(m_A t -
\vec{k}\cdot \vec{x}_i)}.
\end{eqnarray}
We normalize the coupling of the dark photon in terms of the
electromagnetic (EM)  coupling constant $e$.  The ratio of
the dark photon coupling strength to the EM coupling strength is
given by $\epsilon$.
$M_i$ and $q_{D,i}$ are the total mass and dark charge of
the $i$th test object. If the dark photon is a  gauge field associated with baryon number, $U(1)_{B}$,
$q_D$ is the total baryon number; for $U(1)_{B-L}$, $q_{D}$  counts the neutrons in the material.

A GW detector is sensitive to the differential relative displacement between
pairs of test objects along different axes.  This displacement will be induced by slightly
differing forces on the test masses, a difference determined
by the relative phase of the dark photon field at the positions of
the test objects.

For dark photon masses we consider, this phase difference is small
and results in a suppression. The arm lengths are 4 km and
$2.5\times 10^6$ km for LIGO and LISA. For $v_{0}$= $O(10^{-3})$, as
long as the dark photon oscillation frequency is smaller than
${\mathcal O}(10^8)$ Hz  (${\mathcal O}(10^2)$ Hz) for LIGO ( LISA),
the arm length is always much smaller than the wavelength of the
dark photon background.  In contrast, the best sensitivities of
these experiments are at  ${\mathcal O}(10^2)$ Hz  (${\mathcal
O}(10^{-2})$ Hz). Thus $|\vec k\cdot (\vec x_1-\vec x_2)|\ll 1$ is a
good approximation in the frequency regimes with the best
sensitivity in both experiments.

With this approximation, and
noting the test object pairs are composed of the same elements, i.e.
they have the same $\frac{q_{D,i}}{M_i}$, the amplitude of the
induced differential strain in one Michelson interferometer
(relative displacement $\Delta L$ divided by arm length $L$) can be
calculated as
\begin{eqnarray}\label{eq:disp}
R \equiv \frac{\Delta L}{L} \simeq  C\frac{q_{D}}{M} \epsilon e
|\vec A_0| v_{0}.
\end{eqnarray}
Here $C$ is the geometric factor found by averaging over the
direction of DM propagation  and the dark photon polarization,
accounting for the orientation of the GW detector arms. This is
generically ${\mathcal O}(1)$. In the Appendix, we show $C_{\rm
LIGO}=\frac{\sqrt{2}}{3}$ and $C_{\rm LISA}=\frac{1}{\sqrt{6}}$. For
a $U(1)_B$ and $U(1)_{B-L}$ gauge boson acting on a mirror composed
of Silicon, $\frac{q_{D}}{M}\simeq \frac{1}{\ \textrm{GeV}}$ and
$\frac{1}{\ 2 \, \textrm{GeV}}$, respectively. To arrive at
Eq.~(\ref{eq:disp}), we use the instantaneous acceleration of
Eq.~(\ref{eq:acc}), and compute the displacement as a function of
time.

\section{Experimental Sensitivity to a Near-Monochromatic Stochastic GW Background}

While the oscillation frequency of the DPDM field is determined by
the dark photon mass, the virial velocity broadens the oscillation
frequency, i.e. $\Delta f/f\sim v_{0}^2$.  Since $v_0$ is ${\mathcal
O}(10^{-3})$, the signal is nearly monochromatic.

In this section, we begin by examining the experimental sensitivity of
a GW detector to a near-monochromatic stochastic gravitational wave.
We will then rephrase this monochromatic GW sensitivity in terms of
a limit on the dark photon.  We emphasize that this is a
calculational tool; no gravitational waves are present.

A sinusoidal, linearly polarized gravitational plane wave with
frequency $f$ and strain $h(\vec r,t)$, has energy density~\cite{shapiroteukolsky}
\begin{eqnarray}
\rho_{GW}(f)=\frac{\langle\dot h^2\rangle}{16\pi G} = (2\pi
f)^2\frac{\langle h^2\rangle}{16\pi G}.
\end{eqnarray}
Here the average is over time in a local region. For a plane wave
with amplitude $h_0$, $\langle h^2\rangle=\frac{1}{2}h_0^2$. The
one-sided power spectrum of GW strain for such a near-monochromatic
GW can be written in the customary form in terms of the fraction of
the critical density attributable to gravitational waves
\cite{Thrane:2013oya}:
\begin{eqnarray}
S_{GW}(f)=\frac{3 H_0^2}{2\pi^2}f^{-3}\Omega_{GW}(f),
\end{eqnarray}
with
\begin{eqnarray}
\Omega_{GW}(f)\equiv \frac{f}{\rho_c}\frac{d\rho_{GW}}{df} =
\frac{f}{\rho_c} \frac{\Delta\rho_{GW}(f)}{\Delta f},
\label{eqnOmega}
\end{eqnarray}
\noindent where the critical density $\rho_c$ is related to the
Hubble constant $H_0$ as $\rho_c=\frac{3 H_0^2}{8\pi G}$, and we
have
\begin{eqnarray}
S_{GW}(f)=\frac{h_0^2 }{2 \Delta f}.
\end{eqnarray}
In Eq.~(\ref{eqnOmega}), we specialized to the frequency window
$\Delta f$ where the signal (stochastic GW or DPDM) would lie.

The detection of a stochastic cosmological background disturbance with a single detector is difficult, because it may be indistinguishable from other unknown sources of noise.
Using
cross-correlation between comparable, independent interferometers,
however, permits a dramatically better sensitivity via integration
of the correlation over time. To calculate the achievable
signal-noise-ratio (SNR) for a near-monochromatic GW signal, we
follow the analogous SNR calculation for LIGO broadband stochastic
searches based on cross-correlation of GW strain signals between
different interferometers \cite{Abbott:2003hr}.

The expectation value and variance of the standard stochastic GW detection statistic can be written as
\begin{eqnarray}
S&=&\frac{T}{2}\int df\ \gamma(|f|)\ S_{GW}(|f|)\ \tilde
Q(f),\nonumber\\
N^2&=&\frac{T}{4}\int df\ P_1(|f|)\ |\tilde Q(f)|^2\ P_2(|f|).
\end{eqnarray}
The SNR is $S/N$. $T$ is the operation time of the GW experiment,
and we take $T=2$ years. $\gamma(|f|)$ is the overlap reduction
function between two GW detectors \cite{Christensen:1992wi}, e.g.,
the LIGO Hanford and Livingston interferometers. $\tilde Q(f)$ is
the Fourier transform of the optimal filter function, and
$P_{1,2}(f)$ are the one-sided strain noise power spectra of the two
detectors.

For a given signal $S_{GW}(|f|)$, $\tilde Q(f)$ should
take the following form in order to maximize SNR, see
\cite{Allen:1996vm} for a derivation,
\begin{eqnarray}
\tilde Q(f)=\mathcal {N}
\frac{\gamma(|f|)S_{GW}(|f|)}{P_1(|f|)P_2(|f|)}.
\end{eqnarray}
$\mathcal {N}$ is the normalization factor, which will be dropped
when calculating SNR. For a near-monochromatic GW with width
$\Delta f$,
we then find
\begin{eqnarray}
  \textrm{SNR}=\frac{\gamma(|f|) h_0^2 \sqrt{T}}{2\sqrt{P_1(f) P_2(f)\Delta f}}.
  \label{eqnSNR}
\end{eqnarray}

\section{Comparison of a DPDM search with a stochastic GW search}
Interferometer response to a dark photon dark matter field is similar to that to a stochastic gravitational wave background, hence the similarities in analysis methods described in the preceding section. There are some important differences to keep in mind, however. Most important are the inherently long coherence length of the DPDM signal, which ensures a strong simultaneous correlation in the interferometer responses, and the long coherence time ($\sim10^6/f$, or about $10^4$ seconds for a 100-Hz signal), which restricts the bandwidth of the signal, permitting a high signal-to-noise ratio (SNR).

For a stochastic GW signal, the overlap function is ${\mathcal
O}(1)$ at long wavelength and falls off for shorter wavelengths, set
by the separation between the two detectors. $\gamma(|f|)$ falls off
rapidly above $\sim$10 Hz for Hanford and Livingston.
For our signal, the coherence length is enhanced by $1/v$, so the fall off in $\gamma(|f|)$ is unimportant below $\sim 10^{4}$ Hz, well above the best sensitivity.
This implies
$|\gamma|$  near unity for the Hanford and Livingston
interferometers, which are, by design, nearly aligned with each
other, albeit with a rotation by 90$^{\circ}$ that introduces a
relative sign flip in $\Delta L$ and with a misalignment of the
normal vectors to the planes of the interferometers by 27$^{\circ}$.
As a result, the normalized overlap reduction function, averaged
over all directions of the wave vector $\vec k$ and field
polarization $\vec A$, is $-0.9$  for the Hanford and Livingston
interferometers. The overlap reduction functions for the three pairs
of LISA Michelson interferometers are also ${\mathcal O}(1)$, but
instrumental correlations require construction of synthetic
noise-orthogonal interferometers for cross-correlation signal
extraction. We follow the treatment of~\cite{Romano:2016dpx} in
using the ``$<$AE$>$'' correlation for which we estimate a
normalized overlap reduction function of $-0.29$.

For the DPDM signal, the dark matter velocity distribution
would be well modeled as a Maxwell-Boltzmann distribution with a
cutoff at the escape velocity:
\begin{eqnarray}\label{eq:boltz}
f(\vec v) \propto e^{-|v|^2/v_0^2}\Theta(v_{esc}-|v|).
\end{eqnarray}
We take $v_0=230$ km/s, and $v_{esc}=544$ km/sec, the central value
given by the RAVE collaboration \cite{Smith:2006ym}.
In frequency space, the signal will be peaked
around $\omega = \sqrt{m^2 +k^2} \approx m (1 + \frac{v_0^2}{2})$,
with a fall-off controlled by the above distribution. In our
calculations, we choose $v \in \{0.2 v_0, 1.8 v_0\}$ in order to
include $90\%$ of the DM energy density. This implies $\Delta
f/f\simeq 0.95\times 10^{-6}$.


These $\Delta f$ are even smaller than the bandwidths of typical continuous wave sources sought from fast-spinning, non-axisymmetric neutron stars in the galaxy, for which Doppler modulations from the Earth's orbit lead to frequency spreads of ${\pm \mathcal O}(10^{-4})f$ over the course of a year~\cite{Riles:2017evm}.\footnote{In the case of our signal, the Doppler effect leads to a small modulated broadening of ${\mathcal O} (10^{-7})$.} Previous directed GW searches using stochastic analysis techniques, {\it e.g.}, from the Supernova 1987A remnant or from the galactic center, have used coarser $\Delta f$ binning than is necessary in a DPDM search~\cite{O1DirectedStochastic}. Those directed GW searches have benefitted from knowing {\it a priori} the phase difference between a pair of interferometers with respect to a fixed direction on the sky. In the DPDM search that phase difference is nearly zero because of the long coherence length of the field.

For current GW observatories, this paper focuses on how a correlation between the nominally identical and nearly aligned Hanford and Livingston interferometers can be exploited at Advanced LIGO design sensitivities.
The Virgo interferometer operating at design sensitivity would
potentially offer improved sensitivity when used in a network cross
correlation. The gain will be modest, however, because the
intrinsic Virgo sensitivity is expected to be worse than LIGO and
the normalized overlap reduction functions with respect to the LIGO
interferometers are quite low in magnitude ($-0.02$ for
Hanford-Virgo and $-0.25$ for Livingston-Virgo).
Virgo could, however, play a useful role in confirming a statistically
significant outlier found in LIGO analysis; a loud-enough outlier
found in Hanford-Livingston cross-correlation could  be visible
with lower strength in the Livingston-Virgo correlation. In
addition, the Virgo interferometer is different enough in design
from the LIGO interferometers that non-Gaussian,
instrumental spectral lines correlated between Hanford and Livingston, which
are extremely difficult to eliminate entirely, given nominally
identical electronics, are less likely to occur at the same frequencies in Virgo. A
notable example is electrical power mains, which unavoidably
contaminate GW strain data at some level, operate in the U.S. at 60
Hz and in Europe at 50 Hz. A detailed analysis of how to exploit
LIGO / Virgo correlations is beyond the scope of this article.
See~\cite{Aasi:2014zwg} for a network stochastic analysis combining
Initial LIGO and Virgo data.


\section{Results}
For a given choice of SNR, one can estimate the minimal value of
``GW amplitude" $h_0$ detectable by a GW experiment, setting $\Delta f$
as described above for our DPDM signal with long coherence time.
In order to translate the limit on $h_0$ to the expected sensitivity
on the dark photon coupling strength normalized to EM coupling
strength $\epsilon^2=\alpha_{D}/\alpha_{EM}$, we need to compare
$h_0$ with the relative displacement $R$ in Eq. (\ref{eq:disp}).
The passage of a GW planewave with magnitude $h_0$ is
equivalent to a relative displacement with $R = h_0/2$.

We consider both exclusion limits, as well as
discovery potential. In the absence of a signal, DPDM can be constrained. Following convention, we set SNR=2
to set the limit as a function of frequency. A 5$\sigma$ local significance (i.e. after
including a trials factor) is quoted as a benchmark for
discovery. Since our signal is almost-monochromatic, i.e. $\Delta
f/f\sim 10^{-6}$, this is effectively a bump-hunt  in
frequency space, and the trials factor is ${\mathcal O}(10^6)$. We therefore take
SNR~$\approx 7$ for discovery.

\begin{figure}
   \includegraphics[width=0.99\columnwidth]{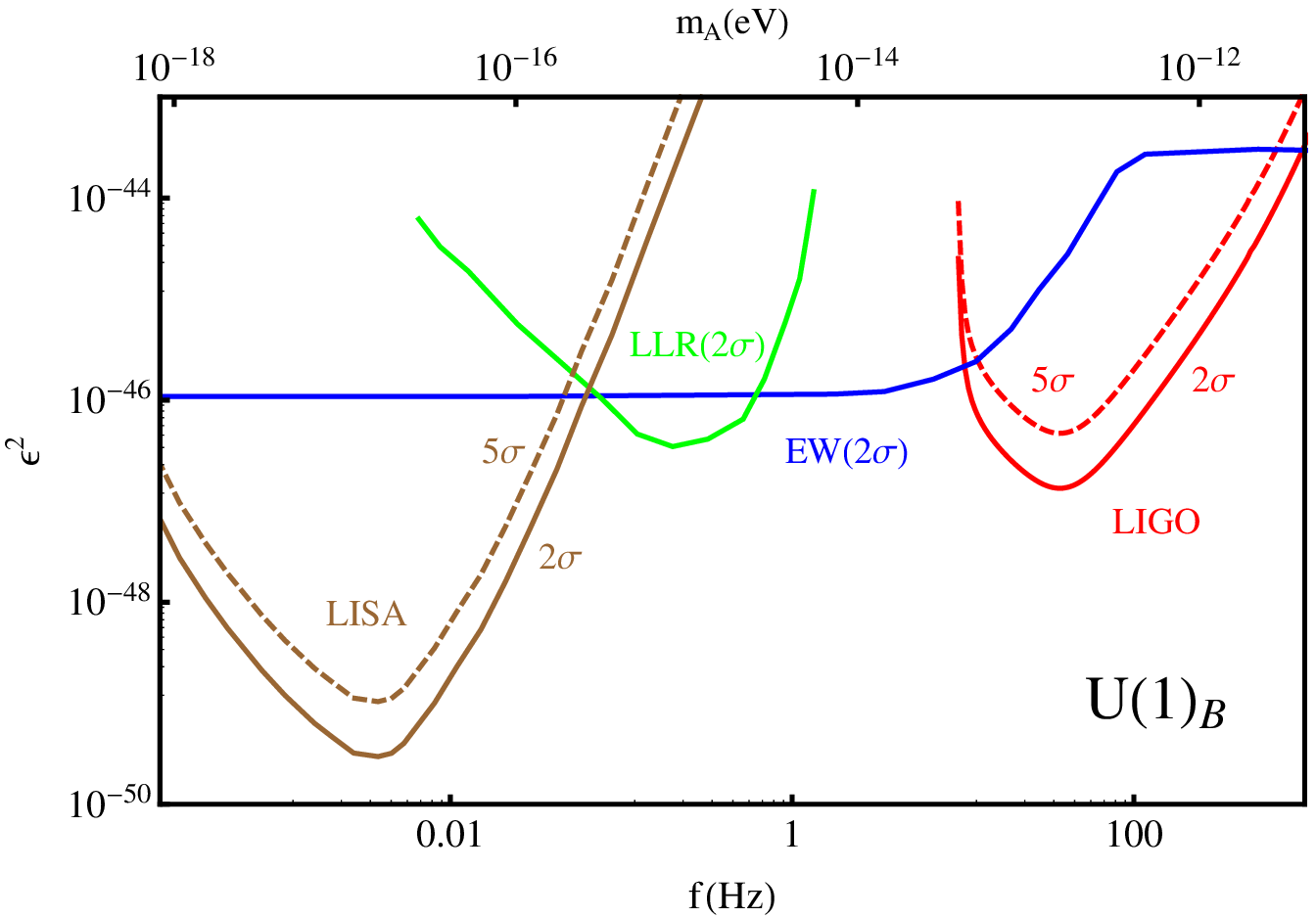}
   \includegraphics[width=0.99\columnwidth]{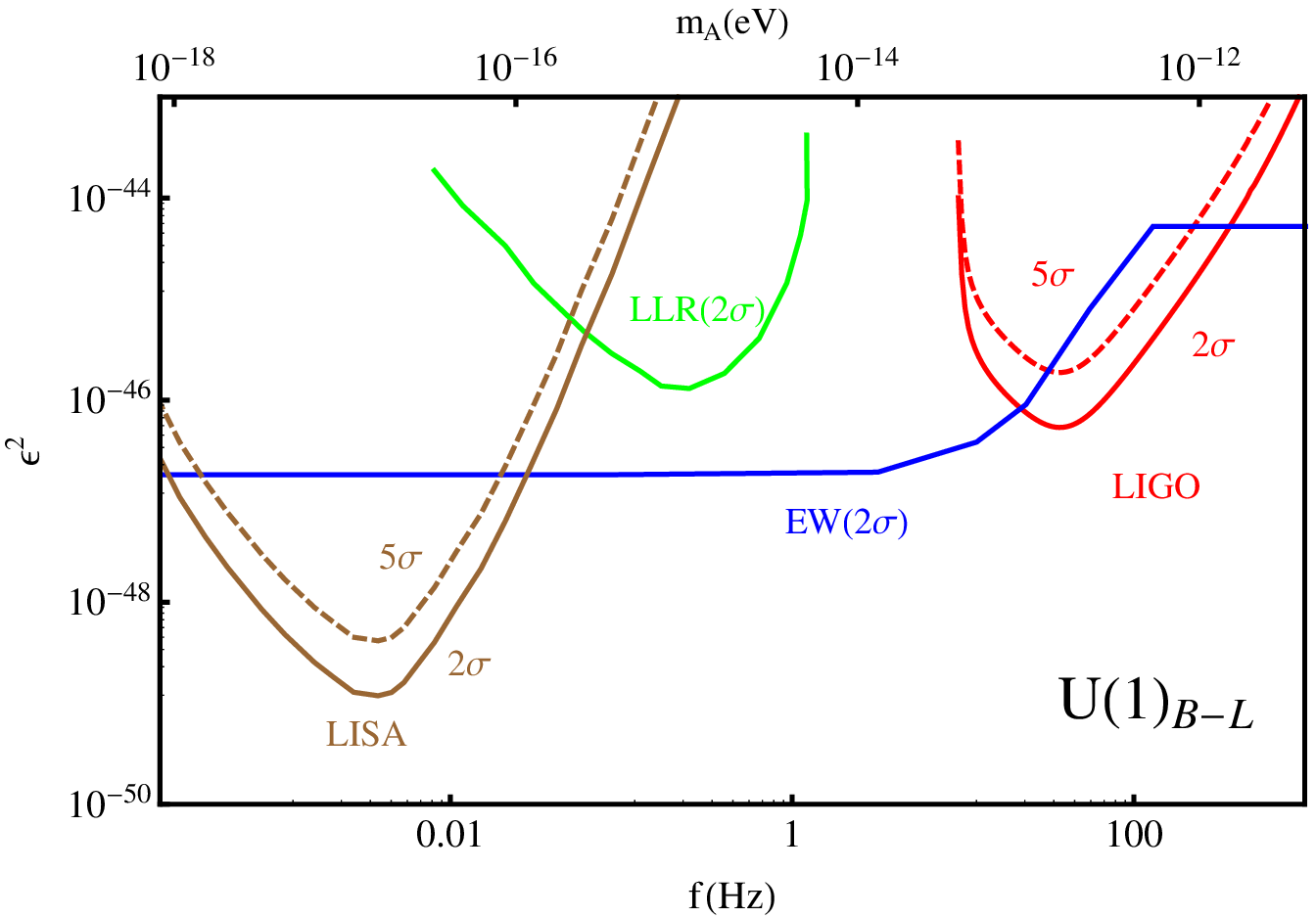}
  \caption{
\label{fig:B-L} The $2\sigma$ exclusion limit and $5\sigma$
discovery potential obtained from LIGO and LISA after 2 years of running for $B$ (upper) and $(B-L)$ (lower) dark photon dark matter.  Coupling strength is normalized to EM coupling strength, i.e. $\epsilon^2=\alpha/\alpha_{EM}$. The
blue and green curves are limits from the E\"ot-Wash (EW) experiment
\cite{Su:1994gu,Schlamminger:2007ht} and the Lunar Laser Ranging
 (LLR) experiment \cite{Williams:2004qba,Talmadge:1988qz,Turyshev:2006gm}.}
\end{figure}

In Fig. \ref{fig:B-L}, we show 2$\sigma$ exclusion limits and
$5\sigma$ discovery potentials in the $\epsilon^2$--frequency plane, assuming the dark
$U(1)$ is the $B$ and $(B-L)$ group, for the LIGO and LISA
experiments. We approximate the LIGO and LISA mirrors as being composed of
silica. $T$ is set to 2 years and $|\gamma(|f|)|$ is chosen to be
0.9 and 0.29 for LIGO and LISA, respectively.
The
one-sided strain noise power spectra for LIGO and LISA are taken
from \cite{Aasi:2013wya,Cornish:2017vip}, with the frequency window
set as described below Eq. (\ref{eq:boltz}).

\section{Conclusion}
We have shown that that GW detectors are potentially sensitive to
the presence of a light gauge field acting as the dark matter.
Present Earth-based interferometers may place the strongest bounds
on $U(1)_{B-L}$ and $U(1)_B$ gauge fields near their peak
sensitivity of ${\mathcal O}(100)$ Hz ($m_{A} \approx 4 \times
10^{-13}$ eV), and in the case of $U(1)_B$, these experiments have
$5\sigma$ discovery potential.  LISA should make comparable progress
in the region of its peak sensitivity, ${\mathcal O}(10^{-2})$ Hz
($m_{A}\approx 4 \times 10^{-17}$ eV). Unlike other bounds on light
gauge fields, these limits are sensitive to the usual astrophysical
uncertainties on the distribution of the dark matter. Variations in
the local dark matter density will directly impact the strength of
the bound, as can variation of the velocity dispersion of the DM,
see Eq.~(\ref{eq:disp}).
For a very-high-SNR detection of DPDM (allowed for LISA and for a 3rd-generation ground-based detector
by current experimental constraints), the signal's spectral line shape would
yield the dark matter speed distribution, and the signal strength's time dependence would
yield directional information, including self-consistency checks.

\begin{acknowledgments}
  {The authors would like to thank M.~Arvanitaki and Y.~Zhong for useful discussions and E.~Thrane for valuable comments on the manuscript.  AP and YZ are supported by the US Department of Energy under grant DE-SC0007859.
  KR is supported by the US National Science Foundation under award NSF-PHY-1505932.}
\end{acknowledgments}

\section{Appendix}
Here, we compute the geometric factor $C$, see Eq.~(\ref{eq:disp}), that characterizes the relative orientations of interferometer arms and the incident dark matter.
Since the dark photon dark matter is non-relativistic, there is no
correlation between the direction of propagation and the
polarization of the gauge field.


For concreteness we will first focus on LIGO where
two arms are orthogonal to each other, and we choose them to be the $x-$ and $y-$axes. The GW detector effectively measures the relative change
of two arm lengths, i.e. $(\Delta L_x - \Delta L_y)$. This can be calculated from Eq.~(\ref{eq:acc}) as
\begin{eqnarray}
&& (\Delta L_x - \Delta L_y) \nonumber\\
&=&\int dt \int dt \{a_x[\cos(m_A t-\vec k\cdot \vec x_1)-\cos(m_A
t-\vec k\cdot \vec
x_2)]\nonumber\\
&-&a_y[\cos(m_A t-\vec k\cdot \vec y_1)-\cos(m_A  t-\vec k\cdot \vec
y_2)]\},
\end{eqnarray}
where $a_x$ and $a_y$ are the accelerations along the $x$ and $y$ axes.
$\vec x_{1,2}$ and $\vec y_{1,2}$ are the position vectors of test
masses and $L$ is the arm length at LIGO: $|\vec{x}_1 - \vec{x}_{2}| = |\vec{y}_1 - \vec{y}_{2}| = L$. Defining the angle between the
wavevector $\vec k$ and the normal to the LIGO plane as $\alpha$, and the angle
between the projected 2D wavevector and the $x$-axis as $\theta$,
the amplitude of the oscillating differential displacement of two
arms is
\begin{eqnarray}
\Delta L&\equiv&|\Delta L_x - \Delta L_y
|_{\textrm{max}}\nonumber\\
&\simeq&|a_x \cos{\theta}-a_y \sin{\theta}| \bigg( \frac{|k| L
\sin{\alpha}}{m_A^2}\bigg).
\end{eqnarray}
We need to perform the average over all possible directions of
$\vec k$ and $\vec a$ (the latter is related to the polarization vector of $A$).   We calculate$\sqrt{\langle \Delta
L^2\rangle}_{\rm LIGO}$, where the $\langle \, \, \rangle$ corresponds to this averaging procedure. This gives
\begin{eqnarray}
\sqrt{\langle \Delta L^2\rangle}_{\rm LIGO} = \frac{\sqrt{2}}{3}
\frac{|a| |k| L}{m_A^2},
\end{eqnarray}
where $a$ is the magnitude of acceleration given in Eq.~(\ref{eq:acc}). The geometric factor of Eq.~(\ref{eq:disp})
is thus $C_{\rm LIGO}=\frac{\sqrt{2}}{3}$.

A similar calculation can be done for LISA where the opening angles
among the three arm pairs are $\pi/3$, giving $C_{\rm LISA} = \frac{1}{\sqrt{6}}$ for
each single interferometer.

From Eq.~(\ref{eq:acc}), and using
$\rho_{DM}\simeq \frac{1}{2}m_A^2 A_{\mu,0}A^{\mu,0}$, we can write
$\Delta L$ in SI units as
\begin{eqnarray}
\sqrt{\langle \Delta L^2\rangle} = C \frac{ \hbar^2 \epsilon e |k|
L}{ m_A^2 c^4 }\frac{q}{M\sqrt{4\pi
\varepsilon_0}}\sqrt{2\rho_{DM}}.
\end{eqnarray}

\end{document}